\documentclass[cameraready]{Interspeech}
\title{A Benchmark for Early-stage Parkinson's Disease Detection from Speech}

\author[affiliation={1}, orcid=0009-0009-2593-5092]{Terry Yi}{Zhong}
\author[affiliation={1}, orcid=0000-0001-5395-0438]{Cristian}{Tejedor-Garcia}
\author[affiliation={1}, orcid=0000-0002-7243-0523]{Khiet P.}{Truong}
\author[affiliation={2}, orcid=0000-0002-0152-9024]{Janna}{Maas}
\author[affiliation={1}, orcid=0000-0001-9964-8760]{Louis}{ten Bosch}
\author[affiliation={2}, orcid=0000-0002-6371-3337]{Bastiaan R.}{Bloem}

\address{
    $^1$ Centre for Language Studies, Radboud University, Nijmegen, The Netherlands \\
    $^2$ Center of Expertise for Parkinson \& Movement Disorders, Department of Neurology, Donders Institute, Radboud University Medical Center, Nijmegen, The Netherlands
}
\email{$^1$\{yi.zhong, cristian.tejedorgarcia, khiet.truong, l.tenbosch\}@ru.nl, $^2$\{janna.maas, bas.bloem\}@radboudumc.nl}
\keywords{Benchmark, Parkinson's disease (PD), Early-stage detection, Speech}

\usepackage[utf8]{inputenc}
\usepackage[T1]{fontenc}
\usepackage{babel}
\usepackage{multirow} % 
\usepackage{hhline} %
\usepackage{subcaption}
\usepackage{hyperref}
\usepackage{seqsplit}
\usepackage{ragged2e}
\usepackage{etoolbox}
\usepackage{enumitem} % For customized lists
\usepackage{multirow}
\usepackage{float}
\usepackage{placeins}
\usepackage{makecell}
\usepackage{graphicx}

\begin{document}

\maketitle

% \raggedbottom

\begin{abstract}
Early-stage Parkinson's disease (EarlyPD) detection from speech is clinically meaningful yet underexplored, and published results are hard to compare because studies differ in datasets, languages, tasks, evaluation protocols, and EarlyPD definitions. To address this issue, we propose the first benchmark for speech-based EarlyPD detection, with a speaker-independent split designed for fair and replicable cross-method evaluation on researcher-accessible datasets. The benchmark covers three common speech tasks and evaluates methods under different training-resource settings. We also present multi-dimensional evaluation breakdowns by dataset, aggregation level, gender, and disease stage to support fine-grained comparisons and clinical adoption. Our results provide a replicable reference and actionable insights, encouraging the adoption of this publicly available benchmark to advance robust and clinically meaningful speech-based EarlyPD detection.
\end{abstract}

%\hspace{2mm}

\section{Introduction}
\label{sec:intro}
Parkinson's disease (PD) is the second most prevalent neurodegenerative disorder, affecting over 10 million people worldwide~\cite{bloem2021parkinson}. Speech impairment can appear early, sometimes years before prominent motor symptoms, and typically worsens with disease progression~\cite{skodda2013progression, smith2018communication}. This has motivated a recent interest in speech-based PD detection as a scalable, non-invasive, and cost-effective approach~\cite{LisannevangelderenInnovativeSpeechBasedDeep2024,review2hossain2025machine,review3sedigh2025voice}.

Most speech-based PD classification studies focus primarily on binary classification between patients with PD and healthy controls (HC), ranging from early feature-based machine-learning (ML) studies~\cite{yuanyuanliuAutomaticAssessmentParkinsons2023,biomarkertelho2024speech} to more recent deep-learning approaches, including self-supervised and pre-trained representations~\cite{BDHPD,InceptionPD-rahmatallah2025pre,zhong2025recapdrobustexplainablecrossattention}. 
%However, in many real-world clinical settings, distinguishing clinically manifest PD from HC is often not a challenge for experienced neurologists, so a generic PD vs.\ HC classifier may offer limited added value. 
Although generic speech-based PD vs.\ HC classifiers provide a valuable foundation, their clinical utility is maximized when they can detect the disease before severe motor symptoms appear. Since neurologists can generally identify clinically manifested PD, reliably detecting \textit{early-stage}\/ PD (referred to as EarlyPD) from speech is highly impactful, as it enables earlier intervention and monitoring~\cite{cao2025speech}.
%More broadly, speech biomarkers may also contribute to differential diagnosis (e.g., atypical Parkinsonism) and longitudinal tracking of disease progression~\cite{MARTINEZMARTIN201550, ParkCeleb_favaroUnveilingEarlySigns2024}.

% - \textbf{Prior studies and limitations}

Nevertheless, speech-based EarlyPD detection, specifically the task of distinguishing EarlyPD from HC, remains comparatively underexplored. First, a large number of papers use the term ``early'' in their work, but do not actually stratify their PD cohorts by disease stage~\cite{early_not2_quamar2025voice,nonEarly2025explainable}. 
Second, even when early-stage groups are defined, cohorts can be highly skewed, limiting representativeness. For instance, one dataset contains almost exclusively male speakers~\cite{ParkCeleb_favaroUnveilingEarlySigns2024}, while another concentrates on PD participants in a single disease stage, as defined by Hoehn and Yahr (H\&Y)~\cite{hoehn1967parkinsonism,early_xvectorjeancolas2021x}. More recently, although still limited, several studies have begun to explore speech-based EarlyPD detection more systematically. 
EarlyPD detection has been performed using classical ML models with handcrafted acoustic features~\cite{early_voicesuppa2022voice}. More recently, multiple ML-based methods have been evaluated across four corpora~\cite{early_ml_2026zebidi2026multilingual}, and deep learning has been explored by stratifying patients using the Movement Disorder Society-sponsored revision of the Unified Parkinson's Disease Rating Scale (MDS-UPDRS)~\cite{UPDRSgoetz2008movement,early_mirco}, achieving strong performance on the picture-description task and taking a step forward.

{However, meaningful comparison across studies remains challenging due to substantial variation in datasets, languages, speech tasks, evaluation protocols, and definitions of EarlyPD.} 
Consequently, these inconsistencies often confound reported results, thereby limiting meaningful comparison of methodological contributions.
 This lack of principled comparability is a central, recognized, and long-standing limitation in speech-based PD detection research.
 
To address this gap, we present the first benchmark for speech-based PD detection, with a specific focus on EarlyPD for clinical relevance. This benchmark establishes a transparent and well-controlled protocol that facilitates fair cross-method comparisons and supports realistic, deployment-oriented clinical assessment. We summarize our main contributions as follows:

\begin{itemize}
  \item We introduce the first benchmark for speech-based EarlyPD detection, enabling fair cross-method evaluation and bridging ML evaluation with realistic usage.
  \item We provide a transparent protocol spanning multiple training conditions and multi-dimensional evaluation. All necessary resources will be made public to ensure reproducibility and facilitate reuse by the research community.
  \item We establish a broad set of baseline results across diverse methods and representative training setups, providing a robust reference and actionable insights for future research.
\end{itemize}

\section{Benchmark Setup}
\subsection{Criteria for Early-Stage PD}\label{section:criteria}

In prior studies, the definition of EarlyPD has not been standardized. Some studies rely on the MDS-UPDRS~\cite{early_mirco}, others use the H\&Y scale~\cite{earlyHYsuppa2022voice}, and many also consider time after diagnosis (TAD), but no consistent rule exists~\cite{early_mirco, TADcutburq2022virtual, rusko2023ewa}. We adopt the eligibility criteria specified in~\cite{TADcutburq2022virtual}: (i) H\&Y stage $\le 2$; (ii) TAD $\le 5$~years. Participants who do not meet these criteria are referred to as the non-Early PD group in this paper.

We prioritize the H\&Y scale for its well-defined cutoff points and straightforward interpretability. While the MDS-UPDRS is effective for measuring clinical outcomes, it lacks widely accepted thresholds to separate disease stages.

% The same MDS-UPDRS total scores can mask heterogeneous profiles, such as pronounced tremor with stability versus severe instability with minimal tremor, where the latter typically indicates more advanced disease.
For disease duration, we adopt a 5~year TAD cutoff, a practical compromise used in~\cite{TADcutburq2022virtual}. While EarlyPD could be defined as untreated (no prior Levodopa), this would be neither clinically representative nor desirable, as many patients initiate treatment soon after diagnosis. Restricting to unmedicated cases would bias the cohort toward atypically mild disease.
\subsection{Open-source Datasets ({Open Track}) }
We screened all potentially accessible PD speech audio datasets, and our findings are highly consistent with a recent survey~\cite{DatasetSurvey-acevedoSurveyOpenVoice2026}. Based on the available documentation and communication with the dataset creators, we found that only PC-GITA~\cite{PCGITAorozco2014new} and NeuroVoz~\cite{mendes2024neurovoz} (hereafter, the benchmark datasets) provide sufficient metadata to identify EarlyPD according to the criteria described in Section~\ref{section:criteria}. 
In short, PC-GITA contains recordings from 100 speakers (50 PD, 50 HC), and NeuroVoz includes 108 speakers (53 PD, 55 HC). Both datasets provide rich clinical metadata (e.g., MDS-UPDRS, H\&Y stage, and TAD), and all PD participants had an official clinical diagnosis. 
While the EWA-DB~\cite{rusko2023ewa} dataset specifies an inclusion criterion of TAD of less than 10 years, it unfortunately does not provide the TAD values needed for reliable filtering, and was hence discarded.

\subsection{Private Dataset (Private Track) } 
Beyond researcher-accessible datasets, we consider a common scenario: many institutions have private datasets that cannot be shared due to regulatory constraints. We therefore keep the main benchmark track fully replicable using accessible data, and provide an additional track that allows incorporating private data to study the benefits of more diverse training resources.

% In this paper, we utilize {ANONYMOUS-DATASET}, a private dataset derived from a single-blind randomized controlled trial conducted in ANONYMOUS-COUNTRY with 200 ANONYMOUS-NATIONALITY speakers with idiopathic PD. The ANONYMOUS %Voice Trainer 
% smartphone app was used to guide participants through the tasks at home, while the speech was recorded using an ANONYMOUS digital voice recorder,
% producing mono WAV audio (44.1kHz, 16-bit). ANONYMOUS-DATASET comprises pre-therapy recordings from participants with sufficient metadata who provided explicit consent for reuse in internal research.

In this paper, we utilize {PERSPECTIVE-Base}, a private dataset derived from the PERSPECTIVE study,
%(“PERsonalized SPEeCh Therapy for actIVE conversation in Parkinson's disease”), 
a single-blind randomized controlled trial conducted in the Netherlands with 200 Dutch speakers with idiopathic PD \cite{maas2022design}. Speech was recorded at participants' homes using the Voice Trainer smartphone app, producing mono WAV audio (44.1kHz, 16-bit) \cite{maas2024effectiveness}. PERSPECTIVE-Base comprises pre-therapy recordings from participants with sufficient metadata who provided explicit consent for reuse in internal research.

\section{Benchmark Protocol}
In this section, we describe our proposed benchmark protocol for {binary} speech-based EarlyPD vs.\ HC detection. We will release all materials needed to {replicate} the benchmark.\footnote{\scriptsize{\url{https://github.com/terryyizhongru/B-EarlyPD-Speech}}}

\subsection{{Task Selection and Configuration}}
\label{taskselection}
All experiments in this paper are trained in a single-task setting. In the open track, we run experiments separately on three speech tasks: sustained vowel, diadochokinetic (DDK), and sentence reading. We only utilize those sub-tasks that are shared across datasets: /a/ for sustained vowel and /pa-ta-ka/ for DDK. When analyzing the impact of adding the private data, we exclude the sentence-reading task to ensure a fair comparison. Our private dataset contains only a single sentence, which would otherwise introduce cross-lingual variation~\cite{ParkCeleb_favaroUnveilingEarlySigns2024}.

\subsection{Early-stage Splits}
\label{splits}
Given the limited number of EarlyPD speakers in the benchmark datasets, we use a fixed, speaker-independent 5-fold split for all experiments. 
To ensure dataset balance, each validation and test set contains 6 EarlyPD and 6 HC speakers. Consequently, 30 EarlyPD speakers are included across the five test folds. The mean age of the EarlyPD group ($66 \pm 9$~years) aligns with the all-stage PD group ($66 \pm 11$~years).
We also ensure that genders are balanced within each dataset and diagnostic group in the validation and test sets. For comparison with the EarlyPD evaluation, we additionally keep a matched HC vs.\ all-stage PD test set (6 PD, 6 HC) from the same sources. All remaining speakers can be used for training.

Since this study focuses on single-task training, we treat recordings from each task as separate training materials. However, the proposed benchmark is not restricted to this setting; future work is encouraged to evaluate multi-task training strategies under the same evaluation protocol.

\subsection{Nested Cross-validation and Evaluation Metrics}\label{sec-nested}
Using the fixed 5-fold splits, we train all models with the same maximum number of epochs and the same early-stopping settings. For each fold, we select the checkpoint with the highest validation area under the receiver operating characteristic curve (AUC). Using this checkpoint, we choose a decision threshold by maximizing the positive-class F1 score on the validation set, and then apply the same threshold to the corresponding held-out test set.
We use AUC and F1 as the main metrics. AUC is used as a threshold-independent metric because it better reflects the discriminative ability of models~\cite{rocferri2011coherent, liAreaROCCurve2024}. F1 is included because it combines recall (sensitivity) and precision, both of which are important in real-world clinical use of speech-based EarlyPD detection.

Each training is run five times with different random seeds, and the average performance and standard deviation (SD) across runs are reported to ensure reliability, following~\cite{zhong2025recapdrobustexplainablecrossattention}.

\begin{table*}[ht!]
\centering
\caption{Results (5 seeds) for all models, tasks, and training data settings. Each entry is Mean~$\pm$~SD of 5 runs, rounded to two decimals. Best F1/AUC within each task block are in bold. Abbreviations: AllPD-sub = AllPD subset, EarlyPD+ = EarlyPD + Private.}
\label{tab:main_table}
\resizebox{\textwidth}{!}{%
\setlength{\tabcolsep}{2.8pt} 
\begin{tabular}{llcccccccccccc}
\toprule
\multirow{2}{*}{\textbf{Model}} & \multirow{2}{*}{\textbf{Metric}}
& \multicolumn{4}{c}{\textbf{DDK}} 
& \multicolumn{4}{c}{\textbf{Vowel}} 
& \multicolumn{3}{c}{\textbf{Sentence}}
& \multirow{2}{*}{\textbf{\textit{Avg}}} \\
\cmidrule(lr){3-6}\cmidrule(lr){7-10}\cmidrule(lr){11-13}
& & \textbf{AllPD} & \textbf{AllPD-sub} & \textbf{EarlyPD} & \textbf{EarlyPD+}
  & \textbf{AllPD} & \textbf{AllPD-sub} & \textbf{EarlyPD} & \textbf{EarlyPD+}
  & \textbf{AllPD} & \textbf{AllPD-sub} & \textbf{EarlyPD} \\
\midrule

\multirow{2}{*}{BDHPD}
& F1  & 0.68$\pm$0.02 & 0.60$\pm$0.06 & 0.66$\pm$0.06 & 0.67$\pm$0.02
      & 0.63$\pm$0.03 & 0.64$\pm$0.03 & \textbf{0.68$\pm$0.04} & 0.66$\pm$0.03
      & 0.70$\pm$0.01 & 0.67$\pm$0.02 & 0.65$\pm$0.02
      & 0.66$\pm$0.03 \\
& AUC & 0.73$\pm$0.04 & 0.65$\pm$0.08 & 0.67$\pm$0.03 & 0.75$\pm$0.03
      & 0.57$\pm$0.04 & 0.50$\pm$0.02 & 0.58$\pm$0.04 & 0.60$\pm$0.04
      & 0.75$\pm$0.02 & 0.70$\pm$0.02 & 0.65$\pm$0.04
      & 0.65$\pm$0.04 \\
\midrule

\multirow{2}{*}{InceptionPD}
& F1  & 0.65$\pm$0.04 & 0.68$\pm$0.04 & 0.65$\pm$0.05 & 0.65$\pm$0.03
      & 0.66$\pm$0.03 & 0.65$\pm$0.02 & 0.66$\pm$0.02 & 0.67$\pm$0.02
      & 0.66$\pm$0.01 & 0.65$\pm$0.01 & 0.64$\pm$0.01
      & 0.66$\pm$0.03 \\
& AUC & 0.69$\pm$0.02 & 0.73$\pm$0.05 & 0.73$\pm$0.03 & 0.71$\pm$0.03
      & 0.61$\pm$0.01 & 0.60$\pm$0.04 & 0.64$\pm$0.05 & \textbf{0.67$\pm$0.03}
      & 0.67$\pm$0.01 & 0.64$\pm$0.02 & 0.61$\pm$0.03
      & 0.66$\pm$0.03 \\
\midrule

\multirow{2}{*}{RECA-PD}
& F1  & \textbf{0.73$\pm$0.04} & 0.65$\pm$0.01 & 0.69$\pm$0.02 & 0.72$\pm$0.04
      & 0.65$\pm$0.05 & 0.64$\pm$0.05 & 0.63$\pm$0.03 & 0.60$\pm$0.05
      & \textbf{0.71$\pm$0.02} & 0.68$\pm$0.03 & 0.68$\pm$0.02
      & \textbf{0.67$\pm$0.03} \\
& AUC & \textbf{0.80$\pm$0.02} & 0.73$\pm$0.01 & 0.77$\pm$0.02 & \textbf{0.80$\pm$0.02}
      & 0.63$\pm$0.05 & 0.59$\pm$0.02 & 0.57$\pm$0.02 & 0.60$\pm$0.03
      & \textbf{0.77$\pm$0.01} & 0.75$\pm$0.02 & 0.73$\pm$0.02
      & \textbf{0.70$\pm$0.02} \\
\midrule

\multirow{2}{*}{\textbf{\textit{Avg}}}
& F1  & \textbf{0.69$\pm$0.03} & 0.64$\pm$0.04 & 0.67$\pm$0.04 & 0.68$\pm$0.03
      & 0.65$\pm$0.04 & 0.64$\pm$0.03 & \textbf{0.66$\pm$0.03} & 0.64$\pm$0.03
      & \textbf{0.69$\pm$0.01} & 0.67$\pm$0.02 & 0.66$\pm$0.02
      & 0.66$\pm$0.03 \\
& AUC & 0.74$\pm$0.03 & 0.70$\pm$0.05 & 0.72$\pm$0.03 & \textbf{0.75$\pm$0.03}
      & 0.60$\pm$0.03 & 0.56$\pm$0.03 & 0.60$\pm$0.04 & \textbf{0.62$\pm$0.03}
      & \textbf{0.73$\pm$0.01} & 0.70$\pm$0.02 & 0.66$\pm$0.03
      & 0.67$\pm$0.03 \\
\bottomrule
\vspace{-0.5cm}
\end{tabular}%
}
\end{table*}

\subsection{Multi-dimensional Evaluation Setup}
We report results separately for each task throughout this paper. Each experiment is first evaluated at the utterance level to reflect model performance. We additionally present aggregate-level performance by pooling a fixed bundle of utterances from the same test speaker (e.g., 3-sustained vowels and 3-sentences). Aggregation is performed by averaging logits (mean-logit) and applying the same decision threshold as in utterance-level evaluation, which better reflects realistic deployment where users provide multiple recordings~\cite{aggregateschwab2019phonemd}. Furthermore, to assess robustness and potential dataset-specific effects, we also stratified performance by dataset and by gender. 
%and encourage future studies to include these breakdowns for more detailed benchmarking.

\section{Experimental Setup}
\subsection{Training Data Settings}
\label{trainingdata}
We benchmark speech-based EarlyPD detection under four training-data settings. To isolate the effect of the PD speakers, we maintain the HC cohorts the same across all configurations:
\begin{enumerate}
  \item \textbf{AllPD (EarlyPD+non-EarlyPD):} Train on the full set of PD speakers across all stages from the benchmark datasets.
  \item \textbf{AllPD-subset:} Train on a distribution-matched subset of AllPD, making the speaker count match that of EarlyPD.
  \item \textbf{EarlyPD:} Train on the full set of EarlyPD speakers from the benchmark datasets.
  \item \textbf{EarlyPD+Private:} Augment EarlyPD with additional EarlyPD speakers from a private dataset, matching the PD training speaker count in AllPD.
\end{enumerate}

{These settings enable three targeted comparisons: (i) comparing Settings 2 and 3 isolates the effect of restricting PD training data to early-stage speakers (at a constant data volume); (ii) comparing Settings 1 and 4 tests whether external EarlyPD data from a private dataset is more beneficial than non-Early PD data within the benchmark datasets; and (iii) comparing Settings 3 and 4 evaluates whether incorporating external EarlyPD speakers improves performance beyond relying solely on accessible EarlyPD data within the benchmark datasets. }

Additionally, we conduct an \emph{all-stage} evaluation under the same AllPD training, where validation and testing are performed on all-stage PD (rather than only EarlyPD), enabling a direct comparison between our EarlyPD benchmark and the conventional all-stage PD detection. All training in this paper uses the same fixed EarlyPD validation and test sets from the benchmark datasets defined in Section~\ref{splits}, except the \emph{all-stage} evaluation, and the training speakers do not overlap with validation/test speakers.

For all speech data, we employ a standard preprocessing pipeline for audio from all datasets: all recordings are converted to mono 16 kHz, 16-bit WAV format and peak-normalized with the SoX toolkit due to the inconsistent recording loudness within and across datasets that we observed.

\subsection{Classification Models}

In this paper, we benchmark EarlyPD detection using three recent open-source speech-based {PD detection} methods representing complementary modeling paradigms:
\begin{itemize}
  \item \textbf{BDHPD}~\cite{BDHPD}: a pretrained self-supervised learning (SSL) speech representation based approach.
  \item \textbf{InceptionPD}~\cite{InceptionPD-rahmatallah2025pre}: a vision pretrained model based approach applied to audio spectrogram ``images''.
  \item \textbf{RECA-PD}~\cite{zhong2025recapdrobustexplainablecrossattention}: an explainable AI speech-based method offering explanations for model decisions.
\end{itemize}
% Together, these methods provide a representative snapshot of how well current advanced deep learning speech-based PD detection methods perform in an early-detection setting.

\subsection{Training Configuration}

All experiments are conducted on an NVIDIA A10 GPU. For all three models, we use their officially released implementations and keep the default hyperparameters~\cite{bdhpd2,pdvoice,recapd}.
%\footnote{\scriptsize
%\url{https://github.com/MorenoLaQuatra/BDHPD};\par
%\ \ \ \url{https://github.com/uams-tri/PD-Voice};\par
%\ \ \ \url{https://github.com/terryyizhongru/RECA-PD}
%}
We apply only minimal and standardized training-configuration adjustments to ensure comparability across models. Specifically, we fix the maximum audio duration to 10s and use consistent FFT parameters to extract spectrograms from 16kHz speech for all tasks. This adjustment is required because the original InceptionPD configuration targets 8kHz audio and 1.5s sustained-vowel recordings, which are inappropriate for DDK and sentence tasks with substantially longer durations. In addition, we standardize the nested cross-validation and checkpoint-selection procedure (Section~\ref{sec-nested}) and adopt a common early-stopping criterion (no improvement for five epochs) with a maximum of 20 training epochs to enable reliable cross-model comparison.

\section{Results {and Discussion}}

% This section examines the benchmark results and discusses their implications for speech-based EarlyPD detection. We first present main benchmark results across training data and model settings (Section~\ref{mainres}), followed by a multi-dimensional evaluation across datasets, aggregation levels, gender, and disease stages (Section~\ref{multiaspect}).

\subsection{Main Results}
\label{mainres}
Table~\ref{tab:main_table} presents the main benchmark results across all training settings, models, and tasks.
{We first examine the results following the comparisons defined in Section~\ref{trainingdata}.
For comparison (i), training exclusively on early-stage patients (EarlyPD) versus the matched subset (AllPD-sub) resulted in improvement on DDK and vowel tasks,} yet sentence reading shows the opposite trend, suggesting that sentence-level cues may benefit from broader disease-stage variability. 
{Across comparisons (ii) and (iii),} we observe consistent gains from adding more PD speakers to training, especially for DDK and sentence tasks, whether the additional data comes from non-Early PD within the benchmark datasets or external EarlyPD cohorts. 
In particular, when comparing with the EarlyPD setting, adding non-Early PD speakers from the benchmark datasets (AllPD) achieves the highest average F1 on the DDK and sentence tasks. Meanwhile, adding external EarlyPD speakers (EarlyPD+Private) generally improves AUC on the shared tasks, sometimes matching and more often exceeding the AllPD setting, while F1 remains comparable. This pattern may reflect the cross-corpus shift and calibration effects: Unlike AUC, F1 relies on a fixed threshold, which may not transfer well when training data are augmented with external speakers.
Overall, these findings suggest that expanding speaker diversity, regardless of whether it originates from external cohorts, is a promising direction.

Across the three models, performance differences appear to vary across speech tasks, suggesting that representation type and training strategy interact with task-specific acoustic and articulatory characteristics. 
In particular, RECA-PD achieves the highest F1 and AUC scores averaged over tasks, with particularly strong performance on DDK and sentence reading. InceptionPD attains competitive AUC values on sustained vowel, while showing comparatively lower performance on the other tasks. BDHPD, originally proposed under a multi-task training regime~\cite{BDHPD}, is evaluated here under a single-task setting to ensure controlled comparison across models.
These findings indicate that explainability-oriented designs can remain competitive within such settings, without implying inherent trade-offs between explainability and predictive performance.

%%%%\cristian{Comparing the three models, RECA-PD achieves the highest mean F1 and AUC scores over all tasks, achieving the top scores in most settings on DDK and sentence reading tasks. InceptionPD often obtain the top AUC on vowels, but is weaker on the other tasks.This pattern suggests that vision-based representations may be less informative for tasks involving more complex articulatory sequences. Note that BDHPD was proposed under a multi-task training regime~\cite{BDHPD}; here we adopt single-task training for controlled comparison across models.These results suggest that an explainability-oriented design need not trade off performance and can even enhance it, further motivating the development of clinically useful XAI methods.}
% Table 3

% Using the average of F1 and AUC, we select the best model–training-data combination for each task. For DDK and Sentences, the best setting is {RECA-PD} trained under the {All-PD} setting. For sustained vowels, the best setting is {InceptionPD} trained under the {EarlyPD + Private} setting.

\subsection{Multi-dimensional Evaluation}
\label{multiaspect}

For a controlled comparison, we report the following multi-dimensional results under the AllPD setting, which performs competitively across most settings, as shown in Table~\ref{tab:main_table}.

First, at the dataset level, Table~\ref{tab:dataset-level} reveals that PC-GITA achieves higher best-case performance than NeuroVoz across tasks, with an average gap of +0.09 in F1 and +0.15 in AUC. Moreover, on most dataset-task pairs, RECA-PD often achieves the highest scores.
{This marked performance difference between the two datasets highlights dataset generalization as a key priority. } Future research should reveal whether this difference is due to linguistic differences or acoustic recording conditions. In particular, the sentence task shows the smallest gap, suggesting comparatively better generalization.

\begin{table}[ht!]
\centering
\vspace{-0.05cm}
\caption{Best per-dataset results (highest mean across models)}
\vspace{-0.1cm}
\label{tab:dataset-level}
\scriptsize
\setlength{\tabcolsep}{4pt}
\begin{tabular}{llcccc}
\toprule
\textbf{Dataset} & \textbf{Metric} & \textbf{DDK} & \textbf{Vowel} & \textbf{Sentence} & \textbf{\textit{Avg}} \\
\midrule
\multirow{3}{*}{PC-GITA}
& Model & RECA-PD & RECA-PD & BDHPD & - \\
& F1    & 0.82$\pm$0.06 & 0.68$\pm$0.04 & 0.73$\pm$0.01 & 0.74$\pm$0.04 \\
& AUC   & 0.91$\pm$0.02 & 0.74$\pm$0.08 & 0.84$\pm$0.04 & 0.83$\pm$0.05 \\
\midrule
\multirow{3}{*}{NeuroVoz}
& Model & RECA-PD & InceptionPD & RECA-PD & - \\
& F1    & 0.63$\pm$0.06 & 0.63$\pm$0.03 & 0.70$\pm$0.01 & 0.65$\pm$0.03 \\
& AUC   & 0.75$\pm$0.03 & 0.53$\pm$0.03 & 0.77$\pm$0.01 & 0.68$\pm$0.02 \\
\bottomrule
\end{tabular}
\vspace{-0.1cm}
\end{table}

%%%%%\textbf{Aggregate-level:}
Second, Table~\ref{tab:aggregate} summarizes the mean deltas between aggregate-level and utterance-level evaluation under AllPD for three aggregation regimes (3 vowels, 3 sentences, and 10 sentences per speaker). Overall, aggregation improves performance as expected, with larger gains in AUC. BDHPD and InceptionPD improve in almost all settings, and the gains generally increase with more aggregated samples. RECA-PD can degrade under small-sample aggregation, but it becomes positive when aggregating 10 sentences. {These results suggest that aggregating multiple recordings effectively mitigates high intra-speaker variability, leading to more robust speaker-level predictions. This underscores the importance of simultaneously evaluating and reporting aggregate-level performance within the proposed benchmark to provide a more comprehensive assessment of model utility in realistic scenarios.}

\begin{table}[ht!]
\centering
\scriptsize
\vspace{-0.2cm}
\caption{Mean $\Delta$ (aggregate-level $-$ utterance-level) results across three aggregation settings}
\vspace{-0.1cm}
\label{tab:aggregate}
\scriptsize
\setlength{\tabcolsep}{6pt}
\begin{tabular}{l p{1.2cm} ccc}
\toprule
\textbf{Model} & \textbf{Metric} & \textbf{3 Vowels} & \textbf{3 Sentences} & \textbf{10 Sentences} \\
\midrule
\multirow{2}{*}{BDHPD}
& $\Delta$F1  & +0.00 & +0.03 & +0.02 \\
& $\Delta$AUC & +0.01 & +0.04 & +0.05 \\
\midrule
\multirow{2}{*}{InceptionPD}
& $\Delta$F1  & +0.01 & +0.03 & +0.00 \\
& $\Delta$AUC & +0.02 & +0.07 & +0.11 \\
\midrule
\multirow{2}{*}{RECA-PD}
& $\Delta$F1  & -0.03 & -0.01 & +0.00 \\
& $\Delta$AUC & -0.03 & +0.02 & +0.05 \\
\bottomrule
\vspace{-0.4cm}
\end{tabular}
\end{table}

%%%%%%\textbf{Gender-level:}
In terms of gender, we observe a consistent trend toward higher performance for female speakers in all models and tasks, shown in the left three columns of Table~\ref{tab:stageandgender}. This contrasts with prior observations that reported higher accuracy on male speakers~\cite{early_xvectorjeancolas2021x,postma25_interspeech}. Since both genders in the EarlyPD test set have similar mean H\&Y stage, TAD, and MDS-UPDRS scores, the reversed trend may reflect dataset variability. {This observed gender disparity motivates further research on fairness and potential dataset biases in speech-based PD detection.}

%%%%%%\textbf{Stage-level:}
Table~\ref{tab:stageandgender} compares performance across disease stages by evaluating the AllPD setting on the all-stage PD and EarlyPD test sets.
Most deltas are positive, indicating that EarlyPD detection is indeed more challenging than all-stage PD detection. A few small negative deltas may reflect training stochasticity or sampling variability in the all-stage test speakers. BDHPD shows a larger gap than the other models, suggesting it may benefit more from all-stage evaluation. 
In addition, the gap between disease stages is largest for the sentence task, consistent with our earlier observation in Section~\ref{mainres} that sentence-level cues may benefit from broader disease-stage variability. 
{The increased difficulty of EarlyPD detection further supports its value as a clinically meaningful setting and may warrant greater attention in future research.}

%% KT testing table with extra columns
\begin{table}[ht!]
\centering
\vspace{-0.2cm}
\caption{Gender columns: Mean~$\Delta$ (female~$-$~male). Disease Stage columns: Mean~$\Delta$ (all-stage~$-$~early-stage). Abbreviations: V = Vowel, S = Sentence}
\vspace{-0.1cm}
\label{tab:stageandgender}
\scriptsize
\setlength{\tabcolsep}{4pt}
\begin{tabular}{lp{0.7cm}|p{0.6cm}p{0.6cm}p{0.6cm}|p{0.6cm}p{0.6cm}p{0.6cm}}
\toprule
& & \multicolumn{3}{c|}{\textbf{Gender}} & \multicolumn{3}{c}{\textbf{Disease Stage}} \\
\textbf{Model} & \textbf{Metric} & \textbf{DDK} & \textbf{V} & \textbf{S} & \textbf{DDK} & \textbf{ V} & \textbf{ S} \\
\midrule

\multirow{2}{*}{BDHPD}
& $\Delta$F1  & +0.07 & +0.05 & +0.05 & +0.04 & +0.05 & +0.07 \\
& $\Delta$AUC & +0.18 & +0.02 & +0.04 & +0.06 & +0.13 & +0.11 \\
\midrule

\multirow{2}{*}{InceptionPD}
& $\Delta$F1  & +0.09 & +0.01 & +0.04 & +0.01 & -0.03 & +0.05 \\
& $\Delta$AUC & +0.13 & +0.09 & +0.09 & +0.03 & -0.02 & +0.10 \\
\midrule

\multirow{2}{*}{RECA-PD}
& $\Delta$F1  & +0.08 & +0.06 & +0.01 & -0.02 & +0.01 & +0.08 \\
& $\Delta$AUC & +0.14 & +0.13 & +0.02 & -0.01 & +0.05 & +0.10 \\
\bottomrule
\vspace{-0.35cm}

\end{tabular}
\end{table}

%%%%\textbf{Task-level:}
 {Finally, zooming into speech tasks, DDK achieves the highest performance across most settings (Tables~\ref{tab:main_table} and \ref{tab:dataset-level}) and exhibits the lowest deltas in the multi-dimensional analysis (Tables~\ref{tab:aggregate} and \ref{tab:stageandgender}), whereas vowel-based evaluation is consistently more challenging, in line with prior findings~\cite{zhong2025recapdrobustexplainablecrossattention,gimeno2025unveiling}. Due to the limited support for spontaneous speech in existing open-source methods, it was not included in the present study. We encourage future work to benchmark spontaneous-speech tasks, which may be particularly informative.}

% \section{Discussion\cristian{ - I recommend to merge Results and Discussion in one single section}}

%  Performance differs markedly between PC-GITA and NeuroVoz, highlighting dataset generalization as a key priority.
% We also observe consistent gains from adding more PD speakers to training, especially for DDK and sentence tasks, whether the additional data come from in-domain later-stage PD or out-of-domain EarlyPD cohorts, suggesting that expanding speaker coverage and diversity remains a promising direction. 
% At the model level, RECA-PD performs best overall, suggesting that explainability-oriented design need not trade off performance and may even enhance it, motivating further development on clinically actionable XAI methods.
% %(while also motivating robustness- and calibration-aware methods under domain shift).
% Task heterogeneity observed in our results should be considered when designing future datasets and benchmarks. Due to limited support in existing open-source methods, we do not include spontaneous speech; we encourage future work to benchmark spontaneous-speech tasks, which may be particularly informative.

% Finally, the observed gender disparity motivates further research on fairness and potential dataset-related biases in speech-based PD detection.

\section{Conclusion}

This paper presents the first benchmark for speech-based EarlyPD detection, addressing the long-standing lack of comparability across prior studies.
This benchmark provides a transparent protocol under different training-resource settings, including open tracks to ensure full comparability and private tracks to study the benefits of more diverse private data.
We publicly release all necessary resources alongside representative baselines, establishing a shared, replicable reference point for future research.
{Our results reveal that while dataset generalization remains a key challenge, expanding speaker diversity is a promising path toward improving EarlyPD detection performance. Meanwhile, the strong performance observed in the explainable model shows that XAI design need not compromise performance, further motivating clinically useful XAI development.}
Additionally, the proposed protocol includes both aggregate-level and gender-stratified results, which are critical for {clinical adoption}.
{We encourage future work to adopt this benchmark as a shared reference point and always report multi-dimensional evaluation results. This collective adoption by the community will ensure transparent and fair cross-method comparisons, ultimately advancing robust and clinically meaningful PD detection from speech.}

\section{Acknowledgments}
% {Anonymous.}
%%%
This publication is part of the project Responsible AI for Voice Diagnostics (RAIVD) with file number NGF.1607.22.013 of the research program NGF AiNed Fellowship Grants, which is financed by the Dutch Research Council (NWO).
This work used the Dutch national e-infrastructure with the support of the SURF Cooperative using grant no. EINF-10519. Finally, we extend our special thanks to the authors of the corresponding data materials for providing us with access.

%%\section{Acknowledgements}
%%Anonymous
%%This work was supported by the NWO research programme AiNed Fellowship Grants under the project Responsible AI for Voice Diagnostics (RAIVD) - NGF.1607.22.013.
 %% PC-GITA
 %%Finally, we extend our special thanksto the authors of the corresponding data materials for providing us with access.
%%Acknowledgement should only be included in the camera-ready version, not in the version submitted for review. The 5th page is reserved exclusively for acknowledgements and  references. No other content must appear on the 5th page. Appendices, if any, must be within the first 4 pages. The acknowledgments and references may start on an earlier page, if there is space.

\section{Generative AI Use Disclosure}
Generative AI tools were used for editing and polishing the language and checking the grammar of this manuscript to improve clarity and readability. The core scientific content, including the proposed benchmark, results, analysis, discussion, and conclusion, was produced solely by the human authors. All authors take full responsibility for the work and its integrity.

\bibliographystyle{IEEEtran}

\bibliography{mybib}

@inproceedings{aggregateschwab2019phonemd,
  title={{PhoneMD: Learning to diagnose Parkinson’s disease from smartphone data}},
  author={Schwab, Patrick and Karlen, Walter},
  booktitle={Proceedings of the AAAI conference on artificial intelligence},
  volume={33},
  number={01},
  pages={1118--1125},
  year={2019}
}

@article{review3sedigh2025voice,
  title={Voice-Based Detection of Parkinson’s Disease Using Machine and Deep Learning Approaches: A Systematic Review},
  author={Sedigh Malekroodi, Hadi and Lee, Byeong-il and Yi, Myunggi},
  journal={Bioengineering},
  volume={12},
  number={11},
  pages={1279},
  year={2025},
  publisher={MDPI}
}

@article{review2hossain2025machine,
  title={Machine learning applications for diagnosing parkinson’s disease via speech, language, and voice changes: A systematic review},
  author={Hossain, Mohammad Amran and Traini, Enea and Amenta, Francesco},
  journal={Inventions},
  volume={10},
  number={4},
  pages={48},
  year={2025},
  publisher={MDPI}
}

@article{smith2018communication,
  title={Communication impairment in Parkinson’s disease: Impact of motor and cognitive symptoms on speech and language},
  author={Smith, Kara M and Caplan, David N},
  journal={Brain and language},
  volume={185},
  pages={38--46},
  year={2018},
  publisher={Elsevier}
}

@article{bloem2021parkinson,
  title={Parkinson's disease},
  author={Bloem, Bastiaan R and Okun, Michael S and Klein, Christine},
  journal={The Lancet},
  volume={397},
  number={10291},
  pages={2284--2303},
  year={2021},
  publisher={Elsevier}
}

@article{early_not2_quamar2025voice,
  title={Voice-based early diagnosis of Parkinson’s disease using spectrogram features and AI models},
  author={Quamar, Danish and Ambeth Kumar, VD and Rizwan, Muhammad and Bagdasar, Ovidiu and Kadar, Manuella},
  journal={Bioengineering},
  volume={12},
  number={10},
  pages={1052},
  year={2025},
  publisher={MDPI}
}

@article{early_ml_2026zebidi2026multilingual,
  title={A multilingual speech analysis framework for robust and explainable early detection of Parkinson’s disease},
  author={Zebidi, Hadjer and BenMessaoud, Zeineb and Frikha, Mondher and Hacine-Gharbi, Abdenour},
  journal={International Journal of Speech Technology},
  volume={29},
  number={1},
  pages={1},
  year={2026},
  publisher={Springer}
}

@article{early_voicesuppa2022voice,
  title={Voice in Parkinson's disease: a machine learning study},
  author={Suppa, Antonio and Costantini, Giovanni and Asci, Francesco and Di Leo, Pietro and Al-Wardat, Mohammad Sami and Di Lazzaro, Giulia and Scalise, Simona and Pisani, Antonio and Saggio, Giovanni},
  journal={Frontiers in neurology},
  volume={13},
  pages={831428},
  year={2022},
  publisher={Frontiers Media SA}
}

@article{nonEarly2025explainable,
  title={Explainable artificial intelligence to diagnose early Parkinson’s disease via voice analysis},
  author={Shen, Matthew and Mortezaagha, Pouria and Rahgozar, Arya},
  journal={Scientific Reports},
  volume={15},
  number={1},
  pages={11687},
  year={2025},
  publisher={Nature Publishing Group UK London}
}

@article{early_xvectorjeancolas2021x,
  title={X-vectors: new quantitative biomarkers for early Parkinson's disease detection from speech},
  author={Jeancolas, Laetitia and Petrovska-Delacr{\'e}taz, Dijana and Mangone, Graziella and Benkelfat, Badr-Eddine and Corvol, Jean-Christophe and Vidailhet, Marie and Leh{\'e}ricy, St{\'e}phane and Benali, Habib},
  journal={Frontiers in Neuroinformatics},
  volume={15},
  pages={578369},
  year={2021},
  publisher={Frontiers Media SA}
}

@article{earlyHYsuppa2022voice,
  title={Voice in Parkinson's disease: a machine learning study},
  author={Suppa, Antonio and Costantini, Giovanni and Asci, Francesco and Di Leo, Pietro and Al-Wardat, Mohammad Sami and Di Lazzaro, Giulia and Scalise, Simona and Pisani, Antonio and Saggio, Giovanni},
  journal={Frontiers in neurology},
  volume={13},
  pages={831428},
  year={2022},
  publisher={Frontiers Media SA}
}

@INPROCEEDINGS{early_mirco,
  author={Plantinga, Peter and Cordelle, Briac and Louër, Dominique and Ravanaelli, Mirco and Klein, Denise},
  booktitle={2025 IEEE 35th International Workshop on Machine Learning for Signal Processing (MLSP)}, 
  title={Does Language Matter for Early Detection of Parkinson's Disease from Speech?}, 
  year={2025},
  volume={},
  number={},
  pages={1-6},
  doi={10.1109/MLSP62443.2025.11204272}}

@article{TADcutburq2022virtual,
  title={Virtual exam for Parkinson’s disease enables frequent and reliable remote measurements of motor function},
  author={Burq, Maximilien and Rainaldi, Erin and Ho, King Chung and Chen, Chen and Bloem, Bastiaan R and Evers, Luc JW and Helmich, Rick C and Myers, Lance and Marks Jr, William J and Kapur, Ritu},
  journal={NPJ digital medicine},
  volume={5},
  number={1},
  pages={65},
  year={2022},
  publisher={Nature Publishing Group UK London}
}

@article{hoehn1967parkinsonism,
  title={Parkinsonism: onset, progression, and mortality},
  author={Hoehn, Margaret M and Yahr, Melvin D},
  journal={Neurology},
  volume={17},
  number={5},
  pages={427--427},
  year={1967},
  publisher={Lippincott Williams \& Wilkins}
}

@article{UPDRSgoetz2008movement,
  title={Movement Disorder Society-sponsored revision of the Unified Parkinson's Disease Rating Scale (MDS-UPDRS): scale presentation and clinimetric testing results},
  author={Goetz, Christopher G and Tilley, Barbara C and Shaftman, Stephanie R and Stebbins, Glenn T and Fahn, Stanley and Martinez-Martin, Pablo and Poewe, Werner and Sampaio, Cristina and Stern, Matthew B and Dodel, Richard and others},
  journal={Movement disorders: official journal of the Movement Disorder Society},
  volume={23},
  number={15},
  pages={2129--2170},
  year={2008},
  publisher={Wiley Online Library}
}

@inproceedings{rocferri2011coherent,
  title={A coherent interpretation of AUC as a measure of aggregated classification performance},
  author={Ferri, Cesar and Hern{\'a}ndez-Orallo, Jos{\'e} and Flach, Peter A},
  booktitle={Proceedings of the 28th International Conference on Machine Learning (ICML-11)},
  pages={657--664},
  year={2011}
}

@article{liAreaROCCurve2024,
  title = {Area under the {{ROC Curve}} Has the Most Consistent Evaluation for Binary Classification},
  author = {Li, Jing},
  editor = {Qin, Hong},
  year = 2024,
  month = dec,
  journal = {PLOS ONE},
  volume = {19},
  number = {12},
  pages = {e0316019},
  issn = {1932-6203},
  doi = {10.1371/journal.pone.0316019},
  urldate = {2025-12-18},
  langid = {english}
}

@article{rusko2023ewa,
  title={{EWA-DB}, Slovak database of speech affected by neurodegenerative diseases},
  author={Rusko, Milan and Sabo, R{\'o}bert and Trnka, Mari{\'a}n and Zimmermann, Alfr{\'e}d and Malaschitz, Richard and Ru{\v{z}}ick{\`y}, Eugen and Brandoburov{\'a}, Petra and Kevick{\'a}, Vikt{\'o}ria and {\v{S}}korv{\'a}nek, Matej},
  journal={medRxiv},
  pages={2023--10},
  year={2023},
  publisher={Cold Spring Harbor Laboratory Press}
}

@article{mendes2024neurovoz,
  title={{NeuroVoz: a Castillian Spanish corpus of parkinsonian speech}},
  author={Mendes-Laureano, Jana{\'\i}na and G{\'o}mez-Garc{\'\i}a, Jorge A and Guerrero-L{\'o}pez, Alejandro and Luque-Buzo, Elisa and Arias-Londo{\~n}o, Juli{\'a}n D and Grandas-P{\'e}rez, Francisco J and Godino-Llorente, Juan I},
  journal={Scientific Data},
  volume={11},
  number={1},
  pages={1367},
  year={2024},
  publisher={Nature Publishing Group UK London}
}

@incollection{DatasetSurvey-acevedoSurveyOpenVoice2026,
  title = {A Survey of Open Voice and Speech Datasets for the Screening and Evaluation of {{Parkinson}}’s {{Disease}}},
  booktitle = {Automatic {{Assessment}} of {{Parkinsonian Speech}}},
  author = {Puerta-Acevedo, Juan C. and Alcalá-Durand, Maria F. and Arias-Londoño, Julián D. and Godino-Llorente, Juan I.},
  year = {2026},
  volume = {2646},
  pages = {31--50},
  publisher = {Springer Nature Switzerland},
  location = {Cham},
  doi = {10.1007/978-3-032-07083-8_3},
  urldate = {2025-12-15},
  isbn = {978-3-032-07082-1 978-3-032-07083-8},
  langid = {english}
}

@article{InceptionPD-rahmatallah2025pre,
  title={Pre-trained convolutional neural networks identify Parkinson’s disease from spectrogram images of voice samples},
  author={Rahmatallah, Yasir and Kemp, Aaron S and Iyer, Anu and Pillai, Lakshmi and Larson-Prior, Linda J and Virmani, Tuhin and Prior, Fred},
  journal={Scientific Reports},
  volume={15},
  number={1},
  pages={7337},
  year={2025},
  publisher={Nature Publishing Group UK London}
}

@article{maas2022design,
  title={Design of the {{PERSPECTIVE}} Study: {{PERsonalized SPEeCh Therapy}} for {{actIVE}} Conversation in {{Parkinson}}’s Disease (Randomized Controlled Trial)},
  author={Maas, Janna J L and De Vries, NM and Bloem, BR and Kalf, JG},
  journal={Trials},
  volume={23},
  number={1},
  pages={274},
  year={2022},
  publisher={Springer}
}

@article{maas2024effectiveness,
  title={Effectiveness of remotely delivered speech therapy in persons with {Parkinson}'s disease--a randomised controlled trial},
  author={Maas, Janna J L and de Vries, Nienke M and IntHout, Joanna and Bloem, Bastiaan R and Kalf, Johanna G},
  journal={EClinicalMedicine},
  volume={76},
  year={2024},
  publisher={Elsevier}
}

@article{cao2025speech,
  title={Speech and language biomarkers for {Parkinson}’s disease prediction, early diagnosis and progression},
  author={Cao, Fangyuan and Vogel, Adam P and Gharahkhani, Puya and Renteria, Miguel E},
  journal={npj Parkinson's Disease},
  volume={11},
  number={1},
  pages={57},
  year={2025},
  publisher={Nature Publishing Group UK London}
}

@inbook{zhong2025recapdrobustexplainablecrossattention,
   title={RECA-PD: A Robust Explainable Cross-Attention Method for Speech-Based Parkinson’s Disease Classification},
   ISBN={9783032025487},
   ISSN={1611-3349},
   url={http://dx.doi.org/10.1007/978-3-032-02548-7_29},
   DOI={10.1007/978-3-032-02548-7_29},
   booktitle={Text, Speech, and Dialogue},
   publisher={Springer Nature Switzerland},
   author={Zhong, Terry Yi and Tejedor-Garcia, Cristian and Larson, Martha and Bloem, Bastiaan R.},
   year={2025},
   month=aug, pages={343–355} }

@INPROCEEDINGS{BDHPD,
  author={La Quatra, Moreno and Orozco-Arroyave, Juan Rafael and Siniscalchi, Marco Sabato},
  booktitle={ICASSP 2025 - 2025 IEEE International Conference on Acoustics, Speech and Signal Processing (ICASSP)}, 
  title={Bilingual Dual-Head Deep Model for Parkinson’s Disease Detection from Speech}, 
  year={2025},
  volume={},
  number={},
  pages={1-5},
  doi={10.1109/ICASSP49660.2025.10889445}}

@article{skodda2013progression,
  title={Progression of Voice and Speech Impairment in the Course of {Parkinson'}s Disease: A Longitudinal Study},
  author={Skodda, Sabine and Gr{\"o}nheit, W and Mancinelli, N and Schlegel, U},
  journal={Parkinson’s Disease},
  volume={2013},
  number={1},
  pages={389195},
  year={2013},
  publisher={Wiley Online Library}
}

@article{LisannevangelderenInnovativeSpeechBasedDeep2024,
  title   = {Innovative speech‐based deep learning approaches for {Parkinson}'s disease classification: A systematic review},
  author  = {van Gelderen, Lisanne and Tejedor-Garcia, Cristian},
  journal = {Applied Sciences},
  volume  = {14},
  number = {17},
  pages   = {7873},
  year    = {2024}
}

@article{yuanyuanliuAutomaticAssessmentParkinsons2023,
  title   = {Automatic assessment of {Parkinson}'s disease using speech representations of phonation and articulation},
  author  = {Liu, Yuanyuan and Reddy, M. Kiran and Penttila, Nelly and Ihalainen, Tiina and Alku, Paavo and Rasanen, Okko},
  journal = {IEEE/ACM Transactions on Audio, Speech, and Language Processing},
  volume  = {31},
  pages   = {242--255},
  year    = {2023}
}

@article{ParkCeleb_favaroUnveilingEarlySigns2024,
  title   = {Unveiling early signs of {Parkinson}'s disease via a longitudinal analysis of celebrity speech recordings},
  author  = {Favaro, Anna and Butala, Ankur and Thebaud, Thomas and Villalba, Jesús and Dehak, Najim and Moro‐Velázquez, Laureano},
  journal = {npj {Parkinson's Disease}},
  volume  = {10},
  number  = {1},
  pages   = {207},
  year    = {2024}
}

@article{biomarkertelho2024speech,
  title   = {Speech as a biomarker for disease detection},
  author  = {Botelho, Catarina and Abad, Alberto and Schultz, Tanja and Trancoso, Isabel},
  journal = {IEEE Access},
  volume  = {12},
  pages   = {184487--184508},
  year    = {2024}
}

@article{gimeno2025unveiling,
  title   = {Unveiling interpretability in self‐supervised speech representations for {Parkinson}'s diagnosis},
  author  = {Gimeno‐G{\'o}mez, David and Botelho, Catarina and Pompili, Anna and Abad, Alberto and Mart{\'\i}nez‐Hinarejos, Carlos‐D.},
  journal = {IEEE Journal of Selected Topics in Signal Processing},
  year    = {2025}
}

@inproceedings{PCGITAorozco2014new,
  title     = {New Spanish speech corpus database for the analysis of people suffering from {Parkinson}'s disease},
  author    = {Orozco‐Arroyave, Juan Rafael and Arias‐Londo{\~n}o, Juli{\'a}n David and Vargas‐Bonilla, Jes{\'u}s Francisco and Gonz{\'a}lez‐R{\'a}tiva, Mar{\'i}a Claudia and N{\"o}th, Elmar},
  booktitle = {Proceedings of the Ninth International Conference on Language Resources and Evaluation ({LREC} 14)},
  pages     = {342--347},
  year      = {2014},
  publisher = {ELRA}
}

@inproceedings{postma25_interspeech,
  title     = {{Evaluating the Effectiveness of Pre-Trained Audio Embeddings for Classification of Parkinson's Disease Speech Data}},
  author    = {Emmy Postma and Cristian Tejedor-Garcia},
  year      = {2025},
  booktitle = {{Interspeech 2025}},
  pages     = {4603--4607},
  doi       = {10.21437/Interspeech.2025-801},
  issn      = {2958-1796},
}

@misc{bdhpd2,
  title        = {{BDHPD Github Repository}},
  author       = {La Quatra, Moreno and Orozco-Arroyave, Juan Rafael and Siniscalchi, Marco Sabato},
  year         = 2025,
  howpublished = {\url{https://github.com/MorenoLaQuatra/BDHPD}},
  note         = {Accessed: 2026-03-04}
}

@misc{pdvoice,
  title        = {{PD-Voice GitHub Repository}},
  author       = {Rahmatallah, Yasir and Kemp, Aaron S. and Iyer, Anu and Pillai, Lakshmi and Larson-Prior, Linda J. and Virmani, Tuhin and Prior, Fred},
  year         = 2025,
  howpublished = {\url{https://github.com/uams-tri/PD-Voice}},
  note         = {Accessed: 2026-03-04}
}

@misc{recapd,
  title        = {{RECA-PD Github Repository}},
  author       = {Zhong, Terry Yi},
  year         = 2025,
  howpublished = {\url{https://github.com/terryyizhongru/RECA-PD}},
  note         = {Accessed: 2026-03-04}
}

\end{document}